\documentclass{PoS}

\title{Recent developments on HYDJET/PYQUEN event generators and novel jet
quenching observables at LHC}

\ShortTitle{Recent developments on HYDJET/PYQUEN event generators and novel jet
quenching observables at LHC}

\author{\speaker{I.P. Lokhtin}, L.V. Malinina, S.V. Petrushanko, 
A.M. Snigirev\\
         D.V. Skobeltsyn Institute of Nuclear Physics, M.V. Lomonosov Moscow
	State University, Moscow, Russia \\
        E-mail: \email{Igor@Lokhtin@cern.ch}}

\author{I. Arsene\thanks{On leave from the Institute for Space Sciences, 
Bucharest, Romania}, K. Tywoniuk\thanks{Curren affiliation: Departamento de 
F{\'\i}sica de Part{\'\i}culas, Universidad de Santiago de Compostela, 
Santiago de Compostela, Spain}\\
       The Department of Physics, University of Oslo, Norway\\}

\abstract{The recent developments on PYQUEN, HYDJET and HYDJET++ event
generators are presented. The partonic energy loss model PYQUEN is implemented
as the modification of the ``standard'' jet event obtained with the generator 
of hadron-hadron interactions PYTHIA. HYDJET and HYDJET++ are the Monte-Carlo event 
generators for the simulation of relativistic heavy ion AA collisions considered as 
a superposition of the soft, hydro-type state and the hard, multi-parton
fragmentation. 
HYDJET++ model is the development and continuation of HYDJET generator, and it
includes more detailed treatment of the ``thermal'' hadronic state generated on 
the chemical and thermal freeze-out hypersurfaces (represented by the 
parameterization of relativistic hydrodynamics with preset freeze-out 
conditions),  collective flow effects and decays of hadronic resonances.  
Some applications of above models to novel jet quenching observables are
discussed.}

\FullConference{High-pT Physics at LHC - Tokaj'08\\
		 March 16 - 19 2008\\
		 Tokaj, Hungary}

\begin{document}
\def\la{\mathrel{\mathpalette\fun <}}
\def\ga{\mathrel{\mathpalette\fun >}}
\def\fun#1#2{\lower3.6pt\vbox{\baselineskip0pt\lineskip.9pt
\ialign{$\mathsurround=0pt#1\hfil##\hfil$\crcr#2\crcr\sim\crcr}}} 
\newcommand{\Pom}{{\hspace{-0.1em}I\hspace{-0.25em}P}}

\section{Introduction}
The experimental and phenomenological study of multi-particle production 
in relativistic heavy ion collisions is expected to provide 
valuable information on the dynamical behaviour of strongly-interacting matter 
in the form of a quark-gluon plasma (QGP), as predicted by 
lattice Quantum Chromodynamics (QCD) calculations. Ongoing and future  
experimental studies in a wide range of heavy ion beam energies require the 
development of new Monte-Carlo (MC) event generators  
and improvement of existing ones. Especially for experiments 
at the CERN Large Hadron Collider (LHC), because of very high parton 
and hadron multiplicities, one needs fast (but realistic) MC tools for heavy 
ion event simulations~\cite{alice1,alice2,cms,Abreu:2007kv}. The main advantage of 
MC technique for the simulation of high-multiplicity hadroproduction is that it allows  
a visual comparison of theory and data, including if necessary the 
detailed detector acceptances, resolutions and responses. The realistic MC 
event generator has to include maximum possible number of observed physical 
effects, which are important for the determination of event topologies: from 
bulk properties of soft hadroproduction (domain of low transverse momenta 
$p_T \la 1$GeV$/c$), such as collective flows, to hard multi-parton production 
in hot and dense QCD-matter, which reveals itself in spectra of high-$p_T$ 
particles and hadronic jets. Moreover, the role of hard and semi-hard particle 
production at LHC can be significant even for bulk properties of created matter, 
and hard probes of QGP became clearly observable in various new 
channels~\cite{Abreu:2007kv}. In most of the available MC 
heavy ion event generators, the simultaneous treatment of collective flow 
effects for soft hadroproduction and hard multi-parton in-medium production 
(medium-induced partonic rescattering and energy loss, so called ``jet
quenching'') is lacking. Thus, in order to analyze existing data on low and 
high-p$_T$ hadron production, to test the sensitivity of physical observables 
at the upcoming LHC experiments (and other future heavy ion facilities) to the QGP 
formation, and study the experimental capabilities of constructed detectors, the 
development of adequate and fast MC models for simultaneous collective 
flow and jet quenching simulations is necessary. HYDJET and HYDJET++ 
event generators includes detailed treatment of hard multi-parton production 
with taking into account known medium effects (such as jet quenching and nuclear 
shadowing), as well as soft hadroproduction.   

\section{PYQUEN event generator}
The event generator for medium-modified nucleon-nucleon collisions  
PYQUEN~\cite{pyquen} is the modification of the ``standard'' jet event 
obtained with the generator of hadron-hadron interactions 
PYTHIA~\cite{pythia}. The details of PYQUEN partonic energy loss model can 
be found in~\cite{Lokhtin:2005px}. The event-by-event simulation procedure 
in PYQUEN includes the following steps. 

\begin{enumerate}
\item Generation of initial partonic state production by calling PYTHIA 
(parton fragmentation being switched off). The spatial vertex of a jet 
production J($x=r\cos{\psi}$,$y=r\sin{\psi}$) is generated according to 
the distribution
\begin{equation} 
\label{vertex}
\frac{dN^{\rm jet}}{d\psi r dr} (b) = \frac{T_A(r_1)\cdot 
T_A(r_2)}{T_{AA}(b)}~, 
\end{equation} 
where $b$ is the impact parameter of a heavy ion collision, $T_A$ and $T_{AA}$ 
are the standard nuclear thickness and nuclear overlap functions, $r_{1,2} 
(b,r,\psi)$ are the distances between the nucleus centers and the vertex J, and $r$ is the distance from the nuclear collision 
axis to vertex J.

\smallskip

\item Then initial PYTHIA-produced partonic state is modified by 
medium-induced radiative and collisional energy loss. For each hard parton 
the following rescattering scheme is applied in the course of a 
proper time $\tau$. 

\smallskip

\noindent
$\bullet$ Calculation of the scattering cross section $\sigma (\tau_i) = 
\int dt~d\sigma/dt$ and generation of the transverse momentum transfer 
$t (\tau_i)$ in the $i$-th scattering in the high-momentum trasfer
limit. \\
$\bullet$ Generation of the displacement between the $i$-th and $(i+1)$-th
scatterings, $l_i = (\tau_{i+1} - \tau_i)$:  
\begin{equation} 
\frac{dP}{dl_i} = \lambda^{-1}(\tau_{i+1}) \exp{(-\int\limits_0^{l_i}
\lambda^{-1} (\tau_i + s)ds)} ~,~~ \lambda^{-1}(\tau ) =\sigma (\tau ) \rho 
(\tau )~,
\end{equation}  
and calculation of the corresponding transverse distance, $l_i p_T/E$. Here 
$\lambda = 1/(\sigma \rho)$ is in-medium mean free path, $\rho \propto T^3$ 
is the medium density at the temperature $T$, $\sigma$ is the integral cross 
section for parton interaction in the medium. \\
$\bullet$ Generation of the energy of a radiated in the $i$-th scattering 
gluon, $\omega _i=\Delta E_{{\rm rad},i}$, according to BDMS radiation 
spectrum~\cite{Baier:1999ds,Baier:2001qw} and its emission angle $\theta _i$ relative 
to the parent parton determined according to the selected by user parameterization. \\
$\bullet$ Calculation of the collisional energy loss in the $i$-th scattering 
\begin{equation} 
\Delta E_{{\rm col},i} = \frac{t_i}{2 m_0}~,
\end{equation} 
where energy of ``thermal'' medium parton $m_0$ is generated according to the 
isotropic Boltzmann distribution at the temperature $T(\tau _i)$. \\
$\bullet$ Reducing the parton energy by collisional and radiative loss per 
each $i$-th scattering,
\begin{equation} 
\Delta E_{{\rm tot},i} = \Delta E_{{\rm col},i} + \Delta E_{{\rm rad},i}~, 
\end{equation}
and changing the parton momentum direction by adding the transverse momentum 
kick due to elastic scattering $i$,
\begin{equation} 
\Delta k_{t,i}^2 =(E-\frac{t_i}{2m_{0i}})^2-(p-\frac{E}{p}\frac{t_i}{2m_{0i}}-
\frac{t_i}{2p})^2-m_p^2~.
\end{equation} 
$\bullet$ Going to the next rescattering, or halting the rescattering if one of the
following two conditions is fulfilled: {\em (a)} the parton escapes the hot QGP 
zone, i.e. the temperature in the next point $T(\tau_{i+1},r_{i+1},\eta_{i+1})$ 
becomes less than $T_c=200$ MeV; or {\em (b)} the parton loses so much energy that 
its transverse momentum $p_T (\tau_{i+1})$ drops below the average transverse 
momentum of the ``thermal'' constituents of the medium, $2T (\tau_{i+1})$.  In 
latter case, such a parton is considered to be ``thermalized" and its momentum in 
the rest frame of the quark-gluon fluid is generated from the random ``thermal'' 
distribution, $dN/d^3p \propto \exp{\left( -E/T\right) }$, boosted to the 
center-of-mass of the nucleus-nucleus collision.

\smallskip

\noindent
\item At the end of each NN collision, adding new (in-medium emitted) gluons to 
the PYTHIA parton list and rearrangement of partons to update string formation 
with the subroutine PYJOIN are performed. An additional gluon is included in the 
same string as its ``parent'', and colour connections of such gluons are
re-ordered by their $z$-coordinates along the string. Then final hadrons are formed 
by calling standard PYTHIA fragmentation routine PYEXEC.   
\end{enumerate}

\section{HYDJET and HYDJET++ event generators} 
HYDJET~\cite{Lokhtin:2005px,hydjet} and HYDJET++~\cite{Lokhtin:2008xi,hydjet++} are 
the Monte-Carlo event generators for the simulation of relativistic heavy ion AA collisions 
considered as a independent superposition of the soft, hydro-type component and the hard, 
multi-parton fragmentation. The hard parts of HYDJET++ and HYDJET are identical, but soft parts are 
different. The treatment of soft component in HYDJET is rather oversimplified. It considers 
pions, kaons and protons/neutrons only, the hadron spectrum being generated as the 
convolution of thermal motion and collective flow~\cite{Lokhtin:2005px}. The soft
part of HYDJET++ is the ``thermal'' hadronic state generated on the chemical and 
thermal freeze-out hypersurfaces represented by the parameterization of relativistic 
hydrodynamics with preset freeze-out conditions(the adapted C++ code 
FAST MC~\cite{Amelin:2006qe,Amelin:2007ic}). It includes longitudinal, radial and 
elliptic flow effects and decays of hadronic resonances. The HYDJET was written in Fortran.  
The main program of HYDJET++ was written in C++ under the ROOT environment~\cite{root} 
(but there is the Fortran-written hard part included in the generator structure as a 
separate directory). 

Before any event generation, HYDJET/HYDJET++ run starts from PYTHIA initialization at 
given c.m.s. energy per nucleon pair $\sqrt{s}$, and then it calculates the total 
inelastic $\sigma _{NN}^{\rm in}(\sqrt{s})$ and hard scattering 
$\sigma_{NN}^{\rm hard}(\sqrt{s},p_T^{\rm min})$ NN cross sections with the minimum 
transverse momentum transfer $p_T^{\rm min}$. If the impact parameter $b$ of heavy ion 
AA collision is not fixed by user, its value $b$ is generated in 
each event between minimum and maximum values in accordance with the differential 
inelastic AA cross section: 
\begin{equation}
\label{sigin_b} 
\frac {d^2 \sigma^{AA}_{\rm 
in}}{d^2b} (b, \sqrt{s}) = \left[ 1 - \left( 1- \frac{1}  
{A^2}T_{AA}(b) \sigma^{\rm in}_{NN} (\sqrt{s}) \right) ^{A^2} \right]~.   
\end{equation} 
If the impact parameter $b$ is fixed, then its value in each event is equal to the 
corresponding input parameter. After specification of $b$ for each given event, mean 
numbers of binary NN sub-collisions $\overline{N_{\rm bin}}$ and nucleons-participants 
$\overline{N_{\rm part}}$ are calculated:   
\begin{eqnarray} 
\label{nbcol}
& & \overline{N_{\rm bin}}(b,\sqrt{s}) = T_{AA}(b) \sigma _{NN}^{\rm in}(\sqrt{s})~, 
\\  
\label{npart}
& & \overline{N_{\rm part}}(b,\sqrt{s}) = \int\limits_0^{2\pi} d\psi 
\int\limits_0^{\infty} rdrT_A(r_1) 
\left[ 1-\exp{\{ \sigma^{\rm in}_{NN}(\sqrt{s}) T_A(r_2)\} } \right]~.  
\end{eqnarray} 
The next step is the simulation of properly particle production in the event. 
The soft, hydro-type state and hard, multi-parton state are simulated 
independently. When the generation of soft and hard components in each event at given 
$b$ is completed, the event record (information about coordinates and momenta of 
primordial particles, decay products of unstable particles and stable particles) is 
formed as the junction of these two independent event outputs. 

\subsection{Generation of hard multi-parton state}

\begin{enumerate}

\item Calculation of the number of NN sub-collisions $N_{AA}^{\rm jet}$ producing hard 
parton-parton scatterings of selected type with $p_T>p_T^{\rm min}$, 
according to the binomial distribution around its mean value (without 
nuclear shadowing correction yet, $S=1$). For this purpose, each from 
$N_{\rm bin}$ sub-collisions is treated by the comparison of random number $\xi_i$ 
generated uniformly in the interval $[0,1]$ with the probability 
$P_{\rm hard}= \sigma_{NN}^{\rm hard}/\sigma _{NN}^{\rm in}$ produce the hard process. 
The $i$-th sub-collision is accepted if $\xi_i<P_{\rm hard}$, and is rejected in the 
opposite case. 

\item Selecting the type of hard NN sub-collision (pp, np or nn) in accordance with 
the phenomenological formula for number of protons $Z$ in the stable nucleus A, 
$Z=A/(1.98+0.015A^{2/3})$.  For this purpose, every from $N_{\rm jet}$ ``successful'' 
sub-collisions is treated by the comparison of two random numbers $\xi^1_i$ and 
$\xi^2_i$ generated uniformly in the interval $[0,1]$ with the probability $Z/A$. 
The proton-proton sub-collision is selected if $\xi^1_i,\xi^2_i<Z/A$,
neutron-neutron sub-collision --- if $\xi^1_i,\xi^2_i>Z/A$, and proton-neutron  
sub-collision --- in other cases.  

\item Generation of multi-parton production in $N_{\rm jet}$ hard NN 
sub-collisions by calling PYQUEN $N_{\rm jet}$ times (see the previous section).  

\item If nuclear shadowing is switched on and beam ions are Pb, Au, Pd or Ca, each hard 
NN sub-collision is treated  by the comparison of random number $\xi_i$ generated 
uniformly in the interval $[0,1]$ with the shadowing factor $S$ taken from the available 
parameterization~\cite{Tywoniuk:2007xy}. It is determined by the type of initially 
scattered hard partons, momentum fractions taken by the partons at the initial 
hard interaction $x_{1,2}$, the square of transverse momentum transfer in the hard 
scattering $Q^2$, and the transverse position of jet production vertex relative to the the 
nucleus centers $r_{1,2}$. The given sub-collision is accepted if $\xi_i<S$, and is 
rejected in the opposite case. 

\item Formation of hadrons for each ``accepted'' hard NN sub-collision with PYTHIA 
(parton fragmentation being switched on), and final junction of $N_{\rm jet}$ sub-events 
to common array using standard PYTHIA event output format. 
\end{enumerate}

\subsection{Generation of soft ``thermal'' state}
The details of physics models used to generate thermal hadronic states in HYDJET and 
HYDJET++ can be found in~\cite{Lokhtin:2005px} and in~\cite{Lokhtin:2008xi} 
respectively. The following MC simulation procedure is applied to generate the ``soft'' 
thermal component in HYDJET++ (in HYDJET only step (4) for pions, kaons and protons/neutrons 
is used). 

\begin{enumerate}

\item Initialization of the chemical freeze-out parameters. It includes the 
calculation of particle number densities. So far, only the stable hadrons and resonances 
consisting of $u$, $d$, $s$ quarks are taken into account from the SHARE particle data 
table~\cite{share}.  

\item Initialization of the thermal freeze-out parameters (if $T^{\rm th}<
T^{\rm ch}$). It includes the calculation of chemical potentials and particle number 
densities. 

\item Calculation of the effective volume of hadron emission region 
$V_{\rm eff}(b)$, the fireball transverse radius $R_f(b)$, the freeze-out proper time 
$\tau _f(b)$ and emission duration $\Delta \tau _f(b)$, and the mean multiplicity of each 
particle species. Then the multiplicity is generated around its mean value according to 
the Poisson distribution.

\item For each hadron the following procedure to generate its four-momentum is 
applied. 

\noindent
$\bullet$  Generation of four-coordinates of a hadron in the fireball rest frame \\
$x^{\mu}=\{\tau \cosh{\eta},~r \cos{\phi},~r \sin{\phi},~\tau \sinh{\eta}\}$ on
each freeze-out hypersurface segment $\tau(r)$ for the element
$d^3\sigma_{\mu}u^{\mu}=d^3\sigma_0^*=n_0^*(r) \mid 1-(d\tau/dr)^2 \mid ^{1/2}\tau
(r)d^2rd\eta$, assuming $n_0^*$ and $\tau$ functions of r (i.e., independent of
$\eta$, $\phi$). It includes sampling uniformly distributed $\phi$ in the interval 
$[0, 2\pi]$, generating $\eta$ according to the Gaussian distribution 
$\exp(-\eta^2/2\eta_{\rm max}^2)$ and $r$ in the interval $[0, R_f(b)]$) using a 
$100$\% efficient procedure similar to the ROOT routine GetRandom().\\
$\bullet$ Calculation of the corresponding collective flow four-velocities.\\
$\bullet$ Generation of the three-momentum of a hadron in the fluid element 
rest frame \\ $p^*\{\sin \theta_p^*cos \phi_p^*,~\sin \theta_p^*sin 
\phi_p^*,~\cos \theta_p^*\}$ according to the equilibrium distribution function \\  
$f_i^{\rm eq}(p^{0*};T,\mu_i)p^{*2}dp^{*}d\cos\theta_p^* d\phi_p^*$ by sampling 
uniformly distributed $\cos\theta_p^*$ in the interval $[-1,1]$ and $\phi_p^*$ in 
the interval $[0, 2\pi]$, and generating $p^*$ using a $100$\% efficient
procedure (similar to ROOT routine GetRandom()).\\
$\bullet$ The standard von Neumann rejection/acceptance procedure to account the 
difference between the true probability 
$W_{\sigma,i}^* d^3\sigma d^3\vec{p}^{*}/p^{0*}$ and the probability 
$n^{0*} f_i^{\rm eq}(p^{0*};T,\mu_i)d^2\vec{r} d\eta d^3\vec{p}^{*}$
corresponding to the previous simulation steps. 
For this purpose, the residual weight is calculated~\cite{Amelin:2006qe}:  
\begin{equation}
\label{wres2}
W_i^{\rm res}=\frac{W_{\sigma,i}^*d^3\sigma}{n^{0*}p^{0*}f_i^{\rm eq}d^2\vec{r} 
d\eta}=\tau\left(1-\frac{\vec{n^*}\vec{p^*}}{n^{0*}p^{0*}}\right)~. 
\end{equation}
Then the simulated hadron four-coordinate and four-momentum is treated by the 
comparison of $W_i^{\rm res}$ with the random number $\xi_i$ generated uniformly 
in the interval $[0,\max(W_i^{\rm res})]$. The $i$-th hadron is accepted if 
$\xi_i<W_i^{\rm res}$, and is rejected in the opposite case (then the generation of
its four-coordinate and four-momentum is tried again).\\ 
$\bullet$ Boost of the hadron four-momentum in the center mass frame of the event 
using the velocity field $\vec{v}(x)$, that is, 
\begin{equation}
\label{M69} 
p^{0} = \gamma(p^{0*} + \vec{v}\vec{p}^{~*})~,~~~~~
\vec{p} = \vec{p}^{~*} + \gamma (1 + \gamma)^{-1} (p^{0*} +p^0 )\vec{v}~.
\end{equation} 
Note that a high generation speed for this algorithm is achieved because of 
almost 100\% generation efficiency due to nearly uniform  
residual weights $W_i^{\rm res}$.

\item Formation of final hadrons from the two- and three-body decays of 
resonances with the random choice of decay channels according to the branching 
ratios taken from SHARE particle data files. The two- and three- particle ``decayer'' 
programs were developed by N.S.~Amelin for FAST MC 
generator~\cite{Amelin:2006qe,Amelin:2007ic}, and have been implemented in HYDJET++.  
\end{enumerate}

\section{Physics applications of PYQUEN and HYDJET/HYDJET++} 

It was demonstrated in~\cite{Lokhtin:2005px} that PYQUEN and HYDJET models are  
capable of reproducing the main features of jet quenching pattern in Au+Au collisions at 
RHIC (high-$p_T$ hadron spectra and the suppression of azimuthal back-to-back 
correlations). HYDJET++ describes well even wider range of various hadronic 
observables measured in heavy ion collisions at RHIC for different centrality sets and  
kinematic ranges: ratio of hadron yields, pseudorapidity and transverse momentum 
spectra (for both low- and high-$p_T$ domains), elliptic flow coefficients 
$v_2(p_T)$, femtoscopic correlations~\cite{Lokhtin:2008xi}. Thus the above generators 
can be applied for the various simulation studies at LHC energies. The set of PYQUEN/HYDJET 
applications to novel jet quenching observables in heavy ion collisions at the LHC was
summarized in~\cite{Abreu:2007kv}. Let us enumerate the main predicted observables 
(all numbers were obtained for central Pb+Pb collisions at $\sqrt{s}=5.5 A$ TeV and
hadronic jets of a cone size $R=0.5$ and transverse energy $E_T>100$ GeV). 

\paragraph{Nuclear modification factor for jets.} The jet nuclear modification factor is 
defined as a ratio of jet yields in $AA$ and pp collisions normalized on the number of 
binary nucleon-nucleon collisions. The predicted by PYQUEN/HYDJET jet suppression factor 
(due to partial gluon bremsstrahlung out of jet cone and collisional loss) is about $2$ 
and almost independent on jet energy. It is clear that the measured jet nuclear 
modification factor will be very sensitive to the fraction of partonic energy loss carried 
out of the jet cone.

\paragraph{Medium-modified jet fragmentation function.} The ``jet fragmentation function'' 
(JFF), $D(z)$, is defined as the probability for a given product of the jet fragmentation 
to carry a fraction $z$ of the jet transverse energy. Significant medium-modified 
softening of the JFF (by a factor of $\sim 4$ and slightly increasing with $z$) was 
obtained with PYQUEN/HYDJET simulations. In addition, the ``anti-correlation'' between 
softening of the JFF and suppression of the absolute jet rates due to partonic energy loss  
out of jet cone (wide-angle gluon bremsstrahlung and collisional loss) is predicted. 

\paragraph{Jets induced by heavy quarks.} The possibility to observe the medium 
effects in the channel with the production of B-jet tagged by an energetic muon was also 
analyzed. A significant softening of the b-jet fragmentation function (measured with the 
energetic muon) due to b-quark energy loss in QGP is predicted.

\paragraph{Azimuthal anisotropy of jet production.}
The azimuthal anisotropy of particle spectrum is characterized by the second
coefficient of the Fourier expansion of particle azimuthal distribution,
elliptic flow coefficient, $v_2$. The non-uniform dependence of medium-induced
partonic energy loss in non-central heavy ion collisions on the parton
azimuthal angle $\varphi$ (with respect to the reaction plane) is mapped
onto the final jet spectra. The predicted by HYDJET/PYQUEN elliptic flow coefficients for 
jets $v_2^{\rm jet}$ increases almost linearly with the growth of $b$ and becomes a 
maximum at $b \sim 1.6 R_A$ (where $R_A$ is the nucleus radius). After that, the 
$v_2^{\rm jet}$ coefficients drop rapidly with increasing $b$. The maximum estimated 
value is $v_2^{\rm jet}\sim 0.04$. 

\paragraph{$P_T$-imbalance in dimuon tagged jet events.} 
An important probe of medium-induced partonic energy loss in ultrarelativistic
heavy ion collisions is production of a single jet opposite to a gauge boson
such as $\gamma^\star$/$Z^0$ decaying into dileptons. The advantage of such
processes is that the mean initial transverse momentum of the hard jet equal to
the mean initial/final transverse momentum of boson, and the energy lost by the
parton can be estimated from the observed $p_T$-imbalance between the leading
particle in a jet and the lepton pair. Despite the fact that the initial distribution 
is smeared and asymmetric due to initial-state gluon radiation, hadronization effects, 
etc., the predicted by HYDJET/PYQUEN additional smearing and the displaced mean and 
maximum values of the $p_{\rm T}$-imbalance due to partonic energy loss can be 
significant. The $p_{\rm T}$-imbalance between the $\mu ^+\mu ^-$ pair and a leading 
particle in a jet is directly related to the absolute value of partonic energy loss, and
almost insensitive to the form of the angular spectrum of the emitted
gluons and to the experimental jet energy resolution.

\section{Summary}
Among other heavy ion event generators, HYDJET and HYDJET++ are concentrated on the 
detailed simulation of jet quenching effect based on the partonic energy loss model PYQUEN, 
and also reproducing the main features of nuclear collective dynamics by fast (but realistic) 
way. The final hadron state in HYDJET/HYDJET++ represents the superposition of two 
independent components: hard multi-parton fragmentation and soft hydro-type part. The hard 
parts of HYDJET and HYDJET++ are identical. The soft part 
of HYDJET++ contains the important additional features as compared with HYDJET (detailed 
treatment of thermal and chemical freeze-out hypersurfaces, resonance decays, etc.)
HYDJET describes well the main high-$p_T$ observables in heavy
ion collisions at RHIC, while HYDJET++ is capable of reproducing also the bulk event
properties  
(such as hadron spectra and ratios, radial and elliptic flow,
femtoscopic momentum correlations). Both generators are promising tools for the various 
simulation studies in heavy ion collisions at the LHC.

\section{Acknowledgments} I.L. wishes to express the gratitude to the organizers of 
the Workshop ``High-pT Physics at LHC'' for the warm welcome and the hospitality. 
This work was supported by Russian Foundation for Basic Research 
(grants No 08-02-91001 and No 08-02-92496) and Grants of President of Russian Federation 
(No 1007.2008.2 and No 1456.2008.2).


\begin{thebibliography}{99}
\bibitem{alice1} F.~Carminati et al. (ALICE Collaboration), 
\emph{J. Phys.} {\bf G 30} (2004) 1517.
\bibitem{alice2} B.~Alessandro et al. (ALICE Collaboration), 
\emph{J. Phys.} {\bf G 32} (2006) 1295.
\bibitem{cms} D.~d'Enterria et al. (CMS Collaboration), 
\emph{J. Phys.} {\bf G 34} (2007) 2307.
\bibitem{Abreu:2007kv} N.~Armesto (ed.) et al., \emph{J. Phys.} {\bf G 35} (2008)
054001. 
\bibitem{pyquen} http://cern.ch/lokhtin/pyquen~.
\bibitem{pythia} T.~Sjostrand, S.~Mrenna and P.~Skands, \emph{JHEP} {\bf 0605} (2006) 
026. 
\bibitem{Lokhtin:2005px} I.P.~Lokhtin and A.M.~Snigirev, \emph{Eur. Phys. J.} 
{\bf C 45} (2006) 211. 
\bibitem{Baier:1999ds}  R.~Baier, Yu. L.~Dokshitzer, A.H.~Mueller and D.~Schiff, 
\emph{Phys. Rev.} {\bf C 60} (1999) 064902. 
\bibitem{Baier:2001qw} R.~Baier, Yu. L.~Dokshitzer, A.H.~Mueller and D.~Schiff, 
\emph{Phys. Rev.} {\bf C 64} (2001) 057902.   
\bibitem{hydjet} http://cern.ch/lokhtin/hydro/hydjet.html~.
\bibitem{Lokhtin:2008xi} I.P. Lokhtin, L.V. Malinina, S.V. Petrushanko, A.M. Snigirev, 
I. Arsene, K. Tywoniuk, {\tt arXiv:0809.2708} 
\bibitem{hydjet++} http://cern.ch/lokhtin/hydjet++~.
\bibitem{Amelin:2006qe}  N.S.~Amelin et al., \emph{Phys. Rev.} {\bf C 74} (2006) 
064901.
\bibitem{Amelin:2007ic}  N.S.~Amelin et al., \emph{Phys. Rev.} {\bf C 77} (2008) 
014903.
\bibitem{root} R. Brun and F. Rademakers, \emph{Nucl. Instrum. Meth.} {\bf A 389} 
(1997) 81; (http://root.cern.ch). 
\bibitem{Tywoniuk:2007xy} K.~Tywoniuk, I.C.~Arsene, L.~Bravina, A.B.~Kaidalov 
and E.~Zabrodin, \emph{Phys. Lett.} {\bf B 657} (2007) 170.
\bibitem{share} G.~Torrieri et al., \emph{Comput. Phys. Coommun.} {\bf 167} 
(2005) 229. 
\end{thebibliography}
\end{document}